\documentclass[]{interact}
\usepackage{xcolor}
\usepackage[natbibapa,nodoi]{apacite}
\usepackage[longnamesfirst,sort]{natbib}
\bibpunct[, ]{(}{)}{;}{a}{,}{,}
\renewcommand\bibfont{\fontsize{10}{12}\selectfont}
\usepackage{amsmath,amssymb,amsfonts}
\usepackage{algorithmic}
\usepackage{graphicx}
\usepackage{textcomp}

  \newtheorem{thm}{Theorem}
 
 \newtheorem{lem}{Lemma}
 \newtheorem{prop}{Proposition}
 \newtheorem{defn}{Definition}
 \newtheorem{rem}{Remark}
 
\begin{document}
\title{Stability of Finite Horizon Optimisation based Control without Terminal Weight}
\author{
\name{Wen-Hua Chen 
\thanks{Contact Wen-Hua Chen. Email:w.chen@lboro.ac.uk }} 
\affil{Department of Aeronautical and Automotive
  Engineering, Loughborough University, Leicestershire, LE11 3TU, U.K.}
  }

\maketitle

\begin{abstract}
This paper presents a stability analysis tool for model predictive control (MPC) where control action is generated by optimising a cost function over a finite horizon. Stability analysis of MPC with a limited horizon but without terminal weight is a well known challenging problem. We define a new value function based on an auxiliary one-step optimisation related to stage cost, namely optimal one-step value function (OSVF). It is shown that a finite horizon MPC can be made to be asymptotically stable if OSVF is a (local) control Lyapunov function (CLF). More specifically, by exploiting the CLF property of OSFV to construct a contractive terminal set, a new stabilising MPC algorithm (CMPC) is proposed. We show that CMPC is recursively feasible and guarantees stability under the condition that OSVF is a CLF. Checking this condition and estimation of the maximal terminal set are discussed. Numerical examples are presented to demonstrate the effectiveness of the proposed stability condition and corresponding CMPC algorithm. 
\end{abstract}


\section{Introduction}
Consider a finite horizon optimisation problem involved in model predictive control (MPC), or also known as receding horizon control (RHC), with the performance index defined as
\begin{equation} \label{eq:initFH}
    J'(x(k), U(k))=\sum_{i=0}^{N} l'(x(k+i),u(k+i))
\end{equation}
where $x(k+i), i=0,\ldots,N$ and $U(k)=[u(k),\ldots,u(k+N)]$ are state and control, respectively. $l'(x,u)$ is the stage cost and  $N$ is the length of the horizon.
Lack of stability of some earlier MPC schemes with a finite horizon (\ref{eq:initFH}) has triggered
extensive research on stability of MPC. Stability analysis of MPC has been an active topic in the last three decades and a rich body of knowledge and understanding about stability and related design parameters of various MPC algorithms has been accumulated; e.g. \cite{RawMay09,lee2011model,GruPan17}. 
There are several approaches in establishing the stability of the above schemes. The most widely used is the terminal weight based framework (e.g. \cite{MayRawRao00,CheBalOre00,fontes2001general}). In this framework, there are three important integrants that are closely coupled, namely, a local control Lyapunov function (CLF) used as a terminal
weight to add in the performance index (\ref{eq:initFH}), a terminal set constraint on the
terminal state, and a terminal control satisfying certain conditions for all state in the terminal set \citep{MayRawRao00}.  However, for complicated systems, it is not easy to find a suitable terminal weight since it not only acts as a CLF for a specific system, but also must link back to the stage cost and the optimal cost function. As pointed out in \cite{grune2008infinite}, adding a terminal weight in the cost function brings a few issues. Among others, performance may be distorted by an inappropriate terminal weight. With the addition of a terminal weight, the cost function now consists of two terms: stage cost within the horizon and a terminal weight. If the terminal weight is much larger than the summed stage cost, the influence of the summed stage cost would be too small during the optimisation. Sometimes a terminal weight may have to be selected quite conservatively to make sure that a terminal weight covers the cost-to-go, particularly for systems whose cost-to-go is difficult to calculate or estimate.

Motivated by the above observation, considerable effort has been made to relax the requirements in the terminal weight based MPC framework in the last two decades.
Broadly speaking, there are two main approaches to addressing the issues caused by using a terminal weight. One is to find a generalised terminal weight that is not necessary to be a CLF if certain property holds, e.g. \cite{jadbabaie2005stability,grimm2005model,GruPan17,faulwasser2018economic}. For instance,  stability for unconstrained discrete-time systems under MPC algorithms was
established in \cite{grimm2005model}, where it does not require the
terminal cost to be a local CLF (could be the zero function) if certain
assumptions are satisfied. The other is to drop terminal weights completely, e.g.  \cite{angeli2016theoretical,grune2013economic,GruPan17,faulwasser2018economic}.  Since the stability results are established based on certain asymptotic property of value functions, this approach normally requires to increase the horizon $N$ in the cost (\ref{eq:initFH}) to be \emph{sufficiently large}, or equality terminal constraints have to be included. Along this line, recently there are interesting development with explicit horizon bounds in \cite{grune2008infinite}) and for discount performance costs  in \cite{granzotto2020finite}.
Table 1 in \cite{faulwasser2018economic} provides a nice summary of the state-of-the art about MPC stability.

This paper is interested in further bridging the gap between MPC theory and practical applications. We are interested in a MPC algorithm with a fixed limited horizon as in the cost (\ref{eq:initFH}). Here a limited horizon implies the horizon is relatively short (for example, for the sake of computational burden). As pointed out in \cite{faulwasser2018economic}, many MPC algorithms widely applied in engineering actually do not have a terminal weight but they are stable; for example, \cite{nilsson2015receding,bostrom2019informative,maffei2017cyber,hu2005receding}. Another evidence is all the case studies in the nonlinear model predictive control toolbox of Matlab don't have a terminal cost since it is in general difficult to find a non-conservative terminal weight for a nonlinear system and the resulted MPC algorithm is actually proven to be stable (so no need to add).   
The lack of theoretical foundation, in particular rigorous stability proofs, is regarded as a drawback of finite horizon MPC particularly for safety critical systems, e.g. \cite{grune2008infinite}. Despite all the existing work; e.g. \cite{grune2008infinite,angeli2016theoretical,grune2013economic,GruPan17,granzotto2020finite}, there is a clear gap between available stability analysis tools and practical applications particularly for MPC with a \emph{limited} horizon (e.g. only several prediction steps). For example, there is a significant interest in developing MPC methods with active learning; see survey paper \citep{Mes2018}. However, MPC with active uncertainty learning is very computationally involved and most of research in this area is on developing suboptimal solutions for this problem with only a \emph{limited} horizon (usual only one or two); e.g. see \cite{chen2021dual}. Focusing on optimisation problems with only a limited look-ahead horizon is a main difference between this work and literature in dealing with finite horizon MPC (e.g. \cite{grune2008infinite,angeli2016theoretical,grune2013economic,GruPan17,granzotto2020finite}). So far, there is little work on stability analysis of MPC with a \emph{limited} horizon (e.g. one or two steps) and without terminal weight. \cite{chen2010stability} reported a stability condition for MPC with any length of horizon including only one-step. 


In summary, this paper is interested in developing stability tools for MPC with the cost function (\ref{eq:initFH}) with a limited number of horizon; that is, $N$ is a small number, or only one or two. This group of MPC algorithms have been widely used in practical applications for a number of reasons; e.g. reduced computational load, or avoid the difficult in finding a suitable terminal weight but so far, there are no existing tools available for this group of MPC algorithms so there is a significant gap between MPC theory and its practical applications. 

To address this challenging problem, we define a new stage cost by reformulating the original stage cost $l'$ in (\ref{eq:initFH}). Optimisation of the modified stage cost gives the same control sequence so MPC behaves as the original one. Based on that, the Optimal one-Step Value Function, OSVF, is defined as the value function yielded by optimising the modified one-step stage cost over admissible control. A main contribution in this paper is that \emph{MPC with a fixed limited horizon as in Eq.(\ref{eq:initFH}) can be made to be stable if OSVF is a CLF} (if it is feasible at the beginning).  More specifically, under this condition, a \emph{contractive} terminal set could be constructed and added to the original optimisation problem with the cost function (\ref{eq:initFH}). We show that the corresponding MPC algorithm, referred as CMPC, is recursively feasible with this new \emph{contractive} terminal constraint and guarantees stability. It is noted that a contractive terminal constraint technique was developed to enlarge the domain of attraction for \emph{terminal weight} stabilising MPC algorithm in \cite{limon2005enlarging}.      

A significant implication of the work in this paper is that if OSVF is a CLF, there is no need to add a terminal weight in the cost function, or extend the horizon significantly for the purpose of stability guarantee.  

This paper is organised as follows. In Section \ref{sec:formation},
a finite horizon MPC problem for constrained nonlinear systems without terminal weight is described. 
New stability condition and corresponding CMPC algorithm with stability guarantee 
are developed in Section~\ref{sec:New}. 
In Section~\ref{sec:OSVF}, the property of OSVF being a CLF is investigated and a procedure to estimate the corresponding maximal terminal set if this property holds is proposed. These results are illustrated by two examples to provide more insight. Simulation study for a numerical example is given in Section~\ref{sec:example}. The original finite horizon MPC is unstable with one step horizon. But after showing its OSVF is actually a CLF, with a minimum change of the algorithm by adding a constructed terminal constraint using the property of OSVF (without adding a terminal weight or increasing the horizon), the finite horizon MPC becomes stable.
Finally this paper ends up with conclusions in Section~\ref{sec:conclusion}.

%

\section{Finite Horizon MPC for Nonlinear Systems} \label{sec:formation}
\label{sec:Preliminary}

Consider a constrained nonlinear system
\begin{equation} \label{eq:sys}
    x(k+1)=f(x(k),u(k)),
\end{equation}
or succinctly  $x^+=f(x,u)$, with the input constraints
\begin{equation} \label{eq:constraintU}
    u \in \mathcal{U} 
\end{equation}
and the state constraints
\begin{equation} \label{eq:constraintX}
    x \in \mathcal{X} \mbox{  and }0 \in \mathcal{X}
\end{equation}

The cost function (\ref{eq:initFH}) is widely used in MPC and other optimisation problems as a standard form. However, it is observed that the cost associated with the initial state $x(k)$ is not changed by the control sequence $U'(k)$, and the optimal control at $k+N$, $u(k+N)$, is also not relevant since it does not change the state sequence within the horizon and so the associated state costs.

Motivated by this intuitive observation, we rewrite the cost function (\ref{eq:initFH}) in the form of a new stage cost $l(x^+,u)$ as
\begin{eqnarray} \label{eq:initFHN}
    J(x(k), U(k|k))  =  \sum_{i=0}^{N-1}
    l(x(k+i+1|k),u(k+i|k)) 
\end{eqnarray}
where 
$U(k|k)=[u(k|k)^T, \ldots, u(k+N-1|k)^T]^T$ denotes calculated control sequence at the time instant $k$  and $x(k+i|k), i=1,\ldots,N$, are corresponding
predicted state sequence in the receding horizon where $(\cdot|k)$ explicitly indicates a variable depending on the measurement at time $k$. Furthermore, $x(k|k)=x(k)$. It was acknowledged that although not all the original stage cost $l'$ could be rewritten in the above form as (\ref{eq:initFHN}), many widely used stage costs could be reformulated in this form. See Remark~\ref{rem:costre} for discussion.  

A classic finite MPC setting is that at each $k$, the online optimisation problem is solved  with the cost (\ref{eq:initFHN}), i.e. 
\begin{equation} \label{eq:modifiedOCP} 
    J_M^*(x(k))= \min_{U(k|k) \in \mathcal{U}} J(x(k),U(k|k)) 
\end{equation} 
subject to system dynamics (\ref{eq:sys}), the control constraint (\ref{eq:constraintU}) and the state constraint (\ref{eq:constraintX}). Assuming the existence of a solution, the optimal control sequence and the corresponding state trajectory are denoted by
$U^*(k|k)$
and 
$x^*(k+1|k), \ldots, x^*(k+N|k)$, respectively.
We use the subscript $*$ to indicate the optimal solution, the optimal control sequence or states under the optimal control sequence. 

Only the first control action is implemented in MPC
\begin{equation}
    u(k)=u^*(k|k)
\end{equation}
and the closed-loop system at time $k+1$ is given by 
\begin{equation}
    x(k+1)=f(x(k),u^*(k|k))
\end{equation}

It shall be highlighted that minimisation of the cost function (\ref{eq:initFH}) or (\ref{eq:initFHN}) actually yields the same control sequence so the resulted MPC algorithms have the same behaviours. 

\begin{rem}
Reformulation of the stage cost plays a significant role in our further development. By doing this, essentially a different Lyapunov function, instead of the optimal value function of minimising the cost (\ref{eq:initFH}) as in the current terminal weight MPC framework, is used in establishing the stability of finite horizon MPC in this paper. But more importantly, the modified stage cost is used to define a new value function (OSVF) and our stability analysis tool is established based on the property of this new function. 

\begin{rem} \label{rem:costre}
The stage cost without state and control cross terms, e.g. in the form of $l'(x,u)=l_x'(x)+l_u'(u)$, can be easily reformulated as $l(x^+,u)$. For a quadratic cost function with a cross coupling term of $x$ and $u$, it can be decoupled by using a simple control variable transform $\hat{u}=u-R^{-1}S x$ where $R$ is the control weight and $S$ the cross term weight. Therefore, for a wide class of stage costs, we are able to reformulate (\ref{eq:initFH}) as in (\ref{eq:initFHN}).   
\end{rem}


\end{rem}

\begin{rem}
    One may argue that following the observation that $u^*(k+N|k)$ is always zero, the associated state related cost at $N$ stage, i.e. $l(x(k+N|k),0)$ can be considered as a terminal weight. Then the current terminal weight based stability conditions could be applied. As shown in \cite{chen2010stability}, this approach does not work since this terminal weight does not cover the cost-to-go as required.       
\end{rem}

We will investigate stability of the above described MPC without resorting to adding a terminal weight. To present our results, the following assumptions commonly used in MPC stability analysis are introduced.  


\begin{itemize}
\item[A1:] $f(\cdot,\cdot)$ and $l(\cdot,\cdot)$ are continuously differentiable. Furthermore, $f(0,0)=0, l(0,0)=0$;

\end{itemize}


\begin{defn} A real-valued scalar function $\phi: R_+ \rightarrow R_+$ belongs to
class $\mathcal{K}$ if it is continuous, strictly increasing and
$\phi(0)=0$. $\phi$ belongs to class $\mathcal{K}_\infty$ if $\phi$ belongs to class $\mathcal{K}$ and is radically unbounded.
\end{defn}






\begin{defn}[Positively invariant set \citep{RawMay09}] A closed set $\mathcal{X}_i$  that
contains the origin as its interior is called as a positively
invariant set for a nonlinear autonomous system $x^+=g(x)$ if
$g(x)$ belongs to $\mathcal{X}_i$ for all $x \in \mathcal{X}_i$.
\end{defn}

\begin{defn}[CLF \citep{RawMay09}] \label{def:denCL} Suppose the set $\mathcal{X}$ and closed set $\Omega$, $\Omega \subset \mathcal{X}$, are control invariant for $x^+=f(x,u), u \in \mathcal{U}$. A function $V: R^n \rightarrow \mathcal{R}_{\ge 0}$ is said to be a CLF in $\mathcal{X}$ for the system $x^+=f(x,u), u \in \mathcal{U}$, and the set $\Omega$ in $\mathcal{X}$ if there exist functions $\beta_i \in \mathcal{K}_\infty, i=1,2,3$, defined on $\mathcal{X}$, such that for any $x \in \Omega \subset \mathcal{X}$, there exists a   control $u \in \mathcal{U}$ such that
\begin{equation} \label{eq:OSVF1}
   \beta_1(|x|) \le V(x) \le \beta_2(|x|) 
\end{equation}
and
\begin{equation} \label{eq:OSVF2}
    V(f(x,u)) -V(x) \le -\beta_3(|x|)
\end{equation}
\end{defn}

\section{Stability for finite horizon MPC without terminal weight and algorithm} \label{sec:New}

\subsection{Finite horizon MPC with stability guarantee} \label{sec:FHMPCS}

To facilitate the discussion, based on the modified stage cost $l$, an auxiliary optimisation problem with one-step stage cost is defined as, for any $x \in \Omega \subset \mathcal{X}$ 
\begin{equation} \label{eq:onestepOCP}
    m(x) \triangleq \min_{u \in \mathcal{U}} l(x^+,u); x^+=f(x,u) \in \Omega
\end{equation}
where $\Omega$ is a compact and closed set. $m(x)$ is referred to as optimal One-Step Value Function (OSVF), which will be used extensively in this paper in establishing stability and developing stability guaranteed MPC algorithms.




Define a sublevel set as
\begin{equation} \label{eq:levelset}
    \Omega(\alpha) \triangleq \left\{ x \in R^n: m(x) \leq \alpha, x \in  \mathcal{X} \right\} 
\end{equation}
for some $\alpha >0 $ where $m(x)$ is defined in Eq.(\ref{eq:onestepOCP}). Furthermore, $\alpha_0$ is used to define the maximal control invariant set.  

We now impose an important condition on OSVF, $m(x)$, for establishing stability of finite horizon MPC. 

\begin{itemize}
\item[A2:] The value function $m(x)$ is a CLF for system (\ref{eq:sys}) with respect to the set $\Omega(\alpha)$ defined in (\ref{eq:levelset}) for any $0<\alpha \le \alpha_o$. Moreover, it also satisfies condition (\ref{eq:OSVF1}) for all $x \in \mathcal{X}$.
\end{itemize}


\begin{rem}
Assumption A2 is related to the system dynamics and the stage cost for online optimisation. Section IV is devoted to investigate the detail about under what conditions this assumption is satisfied.    
\end{rem}

It can be observed that if OSVF $m(x)$ is a CLF as in Definition~\ref{def:denCL}, the set $\Omega(\alpha)$ is well defined with $0 \in  \Omega(\alpha)$ since it is positive definite. Based on Assumption A2, we construct a contractive terminal set and add it as a constraint in the original finite horizon MPC, which forms a new algorithm with stability guarantee as described below.

\textbf{CMPC: MPC with Contractive terminal constraints}
\begin{description}
   \item[Step 1] Define system dynamics (\ref{eq:sys}), constraints $\mathcal{X}$ and $\mathcal{U}$, the length of horizon $N$, the stage cost $l(x^+,u)$ as in the cost function (\ref{eq:initFHN}), initial state $x_0$ and initial $\alpha_0$. Let $k=0$, $x(0)=x_0, \alpha(0)=\alpha_0$. Choose $\delta >0$ as a sufficiently small number.
   
    \item[Step 2] Solve the optimisation problem (\ref{eq:modifiedOCP}) 
subject to system dynamics (\ref{eq:sys}), constraints (\ref{eq:constraintU}) and (\ref{eq:constraintX}), and the terminal constraint
\begin{equation} \label{eq:MPCS-NTC}
    x(k+N|k) \in \Omega(\alpha(k))  \subset \mathcal{X}
\end{equation}

\item[Step 3] Apply $u(k)=u^*(k|k)$. Update the terminal set $\Omega(\alpha(k))$ by the following rule:

\begin{equation} \label{eq:alphak}
\begin{aligned}
      \alpha(k+1)&=\min \{m(x^*(k+N|k)), m(x^*(k+1|k))\}  \\
           &  -\delta\|x^*(k+s|k)\|^2
           \end{aligned}
\end{equation} 
   
where $\delta>0$ is sufficiently small and $s$ denotes the instant where the minimum is achieved in the first term. Terminate the update of the terminal set and let $\alpha(k+1) = 0$ if $\alpha(k+1) \le 0$.

\item[Step 4] Let $k=k+1$ and go to Step 2.
\end{description}


Compared with the original finite horizon MPC, the only change in CMPC is that the terminal constraint (\ref{eq:MPCS-NTC}) is added during the optimisation. This is similar to a standard terminal weight based MPC, but there are two crucial differences. First, there is no need to add a terminal weight in the cost function. The other is that the terminal set is updated in Step 3 at each iteration.

\begin{figure}
\begin{center}
\includegraphics[width=14cm]{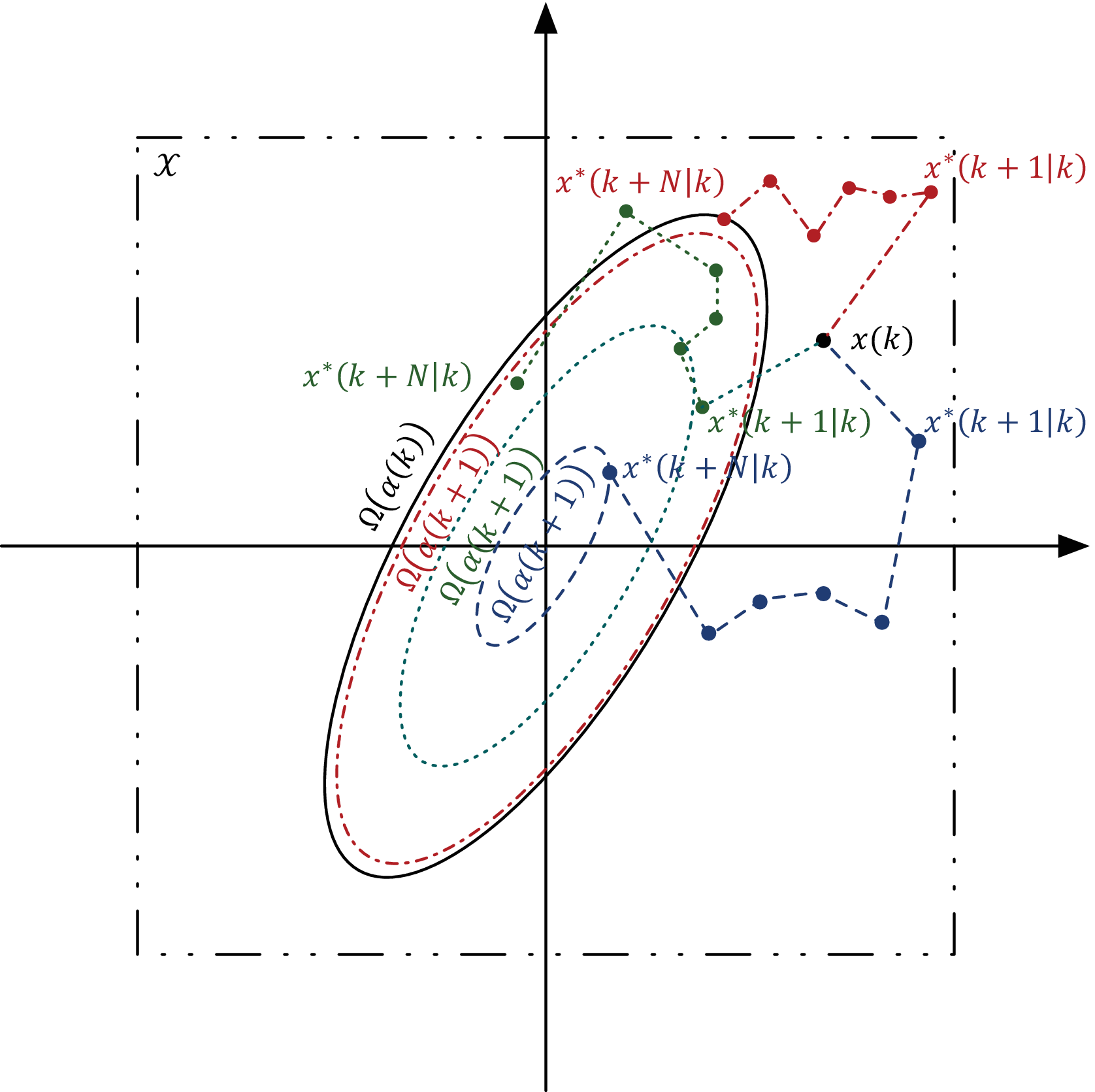}
\caption{Possible optimal trajectories and the corresponding terminal sets when $N=6$. The green dotted  trajectory is for $m(x^*(k+1|k)) \le m(x^*(k+N|k)) \le \alpha(k)$, and both the red dot-dashed and the blue dashed trajectory are for $m(x^*(k+1|k)) > m(x^*(k+N|k))$. The terminal state $m(x^*(k+N|k))$ lies on the boundary of the terminal set $\Omega(\alpha(k))$ for the red dot-dashed while the blue dashed line for the terminal state $m(x^*(k+N|k))$ in side the terminal set $\Omega(\alpha(k))$. The updated terminal sets are depicted by the ellipsoids with corresponding lines.}
\label{fig:alg}
\end{center}
\end{figure}

One distinctive feature of the proposed CMPC is, depending on the relationship between the first state  and the final state in the horizon, we construct the terminal set in a different way. That is, when $m(x^*(k+1|k)) \le m(x^*(k+N|k)) \le \alpha(k)$ as shown by the green dotted trajectory in Figure \ref{fig:alg}, we construct the terminal set $\Omega(\alpha(k+1))$ as the dotted ellipsoid, which allows us to speed up the convergence process of the MPC algorithm by taking advantage of this property. In most of cases, $m(x^*(k+1|k)) > m(x^*(k+N|k))$ as shown in the blue dashed line or the dot-dashed line in Figure~\ref{fig:alg}. The difference between the dashed line and dot-dashed line is that the terminal state $x^*(k+N|k)$ of the red dot-dashed line just lies on the boundary of the terminal set $\Omega(\alpha(k))$ but that of the blue dashed line is inside the terminal set $\Omega(\alpha(k))$. Therefore it is easy to construct a new terminal set $\Omega(\alpha(k+1))$ (blue dashed ellipsoid) using Eq.(\ref{eq:alphak}) for the latter. For the former, it will be shown in Lemma 1 and Theorem 1 that if the OSVF is a CLF, we are able to construct a new contractive terminal set $\Omega(\alpha(k+1))$ using Eq.(\ref{eq:alphak}) (see red dot-dashed ellipsoid in Figure \ref{fig:alg}).           

Theorem~\ref{thm:terminalset} states that the recursive feasibility and asymptotic stability of CMPC can be guaranteed under the condition that OSVF $m(x)$ is a CLF. 

\begin{thm}[Recursive feasibility and asymptotic stability] \label{thm:terminalset}
Consider a nonlinear system~(\ref{eq:sys}) with constraints
(\ref{eq:constraintU}), (\ref{eq:constraintX})  and the performance cost (\ref{eq:initFHN})
without terminal weight, satisfying Assumptions A1-A2. Suppose that CMPC is feasible with respect to an initial state $x_o \in \mathcal{X}$. Then
\begin{enumerate}
    \item CMPC is recursively feasible;
    \item CMPC is asymptotically stable with respect to the origin.
\end{enumerate}
\end{thm}

\begin{rem}
Essentially, Theorem 1 states that a finite horizon MPC is stable if the optimal one-step stage cost is a CLF. Under this condition, a procedure of constructing a feasible terminal set in each iteration is presented. There is no need of significantly modifying the original MPC problem as in other approaches (e.g. adding a terminal weight or significantly increase horizon) in order to guarantee stability. Certainly, not all MPC problems with a finite horizon satisfy this condition although a simple example in Section~\ref{sec:1storder} shows this condition is satisfied for a wide range of MPC settings. If not satisfied, we have to either modify the stage cost $l(x^+,u)$ such that OSVF is a CLF (subject to that satisfactory performance is still achieved), or resort to the existing approaches for stability guarantee; e.g. adding a terminal weight or increasing horizon until the associated conditions are satisfied. The new result is just an addition to the existing tool set for guaranteeing MPC stability.  
\end{rem}
\begin{rem}
The only modification to the original MPC scheme in the proposed approach is to add a (contractive) terminal constraint. Two questions naturally arise: whether or not is it  necessary and how much computation burden would it increase? These two questions have been studied in \cite{mayne2013apologia}. It was concluded that if the system being controlled is subject to hard state constraints and recursive feasibility is desirable, then adding terminal constraints is a necessity. Furthermore, it concludes that \emph{“given that hard state constraints are present, the additional complexity due to the terminal constraint is not significant”}.  In many applications, due to physical constraints, safety or other requirements, hard constraints on state (and control) are present as in the general setting of the MPC problem. These constraints must be respected by not only the terminal states but also all the intermediate states before the terminal state. Hence it is the author’s opinion that this is a \emph{minor} modification to the original MPC algorithm while guaranteeing stability (under the condition that OSVF is a CLF). Furthermore, it shall be highlighted that \emph{CMPC fully recovers to the original one if the terminal constraint is not activated during optimisation}.  
\end{rem}

\subsection{Recursive feasibility and stability proof} 
\begin{lem} \label{lem:J*property}
Suppose that CMPC is feasible for an initial state $ x_0 \in \Omega(\alpha_0)$ for some $\alpha_0$. Under Assumption 2, there exist $\beta_1, \beta_4 \in \mathcal{K}_\infty$ such that its optimal cost $J^*(x)$ satisfying $\beta_1(|x|) \leq J^*(x) \leq \beta_4(|x|)$ for all $x \in \mathcal{X}$.
\end{lem}
\begin{proof}
We establish the lower and upper bounds of the optimal cost $J^*(x)$ under Assumption A2. For any $ x \in \mathcal{X}$, the lower bound is implied by $J^*(x(k)) \ge l(x^*(k+1|k),u^*(k)) \ge m(x(k)) \ge \beta_1(|x(k)|)$. The second inequality is due to the principle of optimality while the last follows from the lower bound of $m(x)$ in A2. The upper bound of $J^*(x)$ can be established using a suboptimal look ahead solution. The optimisation problem (\ref{eq:modifiedOCP}) is split into $N$ one-step optimisation problems and they are solved from Stage $0$ to Stage $N-1$ recursively, which gives a cost $\sum_{i=0}^{N-1} m(x'(k+i|k))$ where $x'(k+i|k)$ is the state under the suboptimal control sequence $u'(k+i|k)$ yielded by this look ahead $N$ strategy. Since $m(x)$ is bounded for any $x \in \mathcal{X}$, the summed stage cost of all the corresponding OSVFs is also bounded.  $J^*(x(k))$ is less than the associated cost of this suboptimal solution so it is also bounded by a function $\beta_4(|x(k)|) \in \mathcal{K}_\infty$. 
\end{proof}

\begin{lem} \label{thm:noT}
Consider a nonlinear system (\ref{eq:sys}) with constraints
(\ref{eq:constraintU}), (\ref{eq:constraintX}), and a terminal constraint (\ref{eq:MPCS-NTC}) and the cost function (\ref{eq:initFHN}) satisfying Assumptions A1-A2. Suppose that the optimisation problem (\ref{eq:modifiedOCP}) starts from a state within an initial feasible set, i.e. $x(0)=x_0 \in \Omega(\alpha_0)$ for a given $\alpha_0$, and there exist a control $u(k+N|k) \in \mathcal{U} $  and a $\mathcal{K}_\infty$ function $\beta_5$ such that   
\begin{eqnarray}    \label{eq:staConNoT}
   & &l(x(k+N+1|k),u(k+N|k))-l(x^*(k+1|k),u^*(k|k))  \nonumber \\  
   && \le  -\beta_5(|x(k)|)
\end{eqnarray} 
for all $k \ge 0$ where $x(k+N+1|k)=f(x^*(k+N|k),u(k+N|k)) \in \Omega(\alpha(k)) $ and $x^*(k+N|k)$ is the terminal state under the optimal control sequence at time $k$. The optimisation problem (\ref{eq:modifiedOCP}) is recursively feasible and the closed-loop system stemming from the MPC algorithm is asymptotically stable.
\end{lem}

\begin{proof} Suppose that the optimisation problem (\ref{eq:modifiedOCP}) is feasible at time $k$. The optimal control sequence, the optimal state trajectory and the optimal cost are denoted by $u^*(k+i|k)$ and $x^*(k+i+1|k)$, $i=0, \ldots,N-1$, and  $J^*(x(k))$, respectively. Following Lemma~\ref{lem:J*property}, $J^*(x(k))$ could be used as a Lyapunov function candidate. 
Stability is established by showing that the modified value function $J^*(x)$ has a monotonicity property, that is 
\begin{equation} \label{eq:monoto}
  J^*(x(k+1))-J^*(x(k))  \le -\beta_5(|x(k)|)
\end{equation}
where $x(k+1)$ denotes the system state at time $k+1$ under
$u(k)=u^*(k|k)$.

Rewrite the optimal cost $J^*(x(k))$ as
\begin{eqnarray} \label{eq:ineq2}
    & & J^*(x(k)) \nonumber \\ 
    & = & l(x^*(k+1|k),u^*(k|k)) \nonumber \\
&&+\sum_{i=1}^{N-1}
    l(x^*(k+i+1|k), u^*(k+i|k)) \nonumber \\
    & &+ l(x(k+N+1|k),u(x+N|k))\nonumber \\
    &&- l(x(k+N+1|k),u(x+N|k))
\end{eqnarray}
with $x(k+N+1|k)=f(x^*(k+N|k),u(k+N|k))$ for any $u(k+N|k) \in \mathcal{U}$.
Following condition (\ref{eq:staConNoT}), under the optimal control sequence $U^*(k|k)$ and corresponding state trajectory, there exist a control $\widetilde{u}(k+N|k) \in \mathcal{U}$ and corresponding $\widetilde{x}(k+N+1|k)=f(x^*(k+N|k),\widetilde{u}(k+N|k)) \in \Omega(\alpha(k))$ such that it holds. Invoking this condition into (\ref{eq:ineq2}) yields
\begin{eqnarray}  \label{eq:ineq1}
 & &   J^*(x(k)) \nonumber \\ & \ge & 
 \sum_{i=1}^{N-1}
    l(x^*(k+i+1|k), u^*(k+i|k))+ \beta_5(|x(k)|) \nonumber \\
    &&+ l(\widetilde{x}(k+N+1|k),\widetilde{u}(k+N|k)) 
\end{eqnarray} 

 The control sequence $U(k+1|k+1) \triangleq [u^*(k+1|k), \ldots, u^*(k+N-1|k), \widetilde{u}(k+N|k)] \in \mathcal{U}$ ensures all the constraints are met and $\widetilde{x}(k+N+1|k) \in \Omega(\alpha(k))$. With $x(k+1)=x^*(k+1|k)$ under $u(k)=u^*(k|k)$, it forms a feasible control sequence for the optimisation problem (\ref{eq:modifiedOCP}) at time $k+1$. Hence it is recursively feasibile.
 
Combining the principle of optimality at time $k+1$ and condition (\ref{eq:ineq1}) obtains Eq.(\ref{eq:monoto}).
Hence the monotonicity of the modified value function is established. Asymptotic stability can be established by following the standard argument in classic stability guaranteed MPC, e.g. \cite{MayRawRao00,RawMay09}. 
\end{proof}

We are now ready to establish Theorem 1. 

\begin{proof}[Theorem 1]
The proof consists of two steps. We first show the recursive feasibility of CMPC under the condition that OSVF is a CLF and then establish its stability.

\emph{Item 1)}. Suppose that at time $k$, with the state $x(k)$, the optimisation problem (\ref{eq:modifiedOCP}) with a terminal set $\Omega(\alpha(k))$ is feasible and solved, which gives the optimal control sequence and corresponding states. The terminal state $x^*(k+N|k)$  belongs to the control invariant set  $\Omega(\alpha(k))$. We now update the terminal set $\Omega(\alpha(k))$ as in Step 3 by Eq.(\ref{eq:alphak}) and show the MPC algorithm is still feasible at $k+1$ with the updated terminal constraint.

To simplify the discussion, let $\alpha'(k+1)=m(x^*(k+s|k))=\min\{m(x^*(k+N|k)), m(x^*(k+1|k))\} $ where $s$ could be either $1$ or $N$. We need to consider two cases, depending on the relationship between $x^*(k+1|k)$ and $x^*(k+N|k)$. First consider the case of $m(x^*(k+1|k)) \ge m(x^*(k+N|k))$. 
In this case, the minimum is achieved at $x^*(k+N|k)$, i.e. $s=N$.
It is straightforward to show that $x^*(k+N|k) \in \Omega(\alpha'(k+1)) \subseteq \Omega(\alpha(k)) $. Following Assumption A2 that $m(x)$ is a CLF with respect to all $x$ in $\Omega(\alpha(k))$, we have that there exists a control $u(k+N|k) \in \mathcal{U}$ such that $x(k+N+1)=f(x^*(k+N|k),u(k+N|k)) \in \Omega(\alpha'(k+1))$. 


Now consider the case that $m(x^*(k+1|k)) <  m(x^*(k+N|k))$, i.e. $s=1$.  First, one can conclude that $\alpha'(k+1) \le \alpha(k)$ since $m(x^*(k+1|k)) \le m(x^*(k+N|k)) \le \alpha(k)$. It follows from $m(x)$ being a local CLF that $\Omega(\alpha'(k+1))$ is a control invariant set. Hence, there always exists a control sequence $\widetilde{u}(k+1|k+1), \ldots,\widetilde{u}(k+N|k+1) \in \mathcal{U}$ such that all states $\widetilde{x}(k+i|k+1), i=2,\cdots,N+1$, including the terminal state $\widetilde{x}(k+N+1|k+1)$ within the constraint set $\Omega(\alpha'(k+1)) \subseteq \Omega(\alpha(k)) \subset \mathcal{X}$. 

Therefore, one has $\alpha'(k+1) \le \alpha(k)$ for either $s=1$ or $s=N$. We now show that for either case,
there always exists a sufficiently small $\delta>0$ such that the update law (\ref{eq:alphak}) holds. This is due to the fact that $\Omega(\alpha'(k+1))$ is a control invariant set. Considering the case of $s=1$, 
there exist $u(k+N|k+1)$ and $\beta_3(\cdot)$ such that
\begin{eqnarray}\label{eq:alphaD}
  &&  m(f(\widetilde{x}(k+N+1|k+1),u(k+N|k+1))  \nonumber\\
     & \le &  m(\widetilde{x}(k+N|k+1)) -\beta_3(|\widetilde{x}(k+N|k+1)|) \nonumber\\ & \le & \alpha'(k+1)-\beta_3(|\widetilde{x}(k+N|k+1)|)
\end{eqnarray} 

Eq.(\ref{eq:alphaD}) implies that there always exists an sufficiently small $0<\delta \|\widetilde{x}(k+N|k+1)\| \le \beta_3(|\widetilde{x}(k+N|k+1)|)$ and a control $\widetilde{u}(k+N|k+1) \in \mathcal{U}$ such that $\widetilde{x}(k+N+1|k+1)=f(\widetilde{x}(k+N|k+1),\widetilde{u}(k+N|k+1))$ falls into the updated terminal set $\Omega(\alpha(k+1))$ with $\alpha(k+1)=\alpha'(k+1)-\delta\|\widetilde{x}(k+N|k+1)\|^2 $ as in (\ref{eq:alphak}). We are able to draw the same conclusion for $s=N$.



In summary, CMPC at $k+1$ is feasible with respect to the constructed terminal set $\Omega(\alpha(k+1))$ if it is feasible at $k$ with respect to a terminal set $\Omega(\alpha(k))$. Furthermore,  one has $\alpha(k+1) < \alpha(k)$ with the update rule of the terminal set in Step 3. When the terminal set shrinks to the origin with $\alpha(k)$ approaching to zero, the terminal set is not updated anymore and the optimisation reduces to one with an equality constraint on the terminal state. By repeating the above process, the recursive feasibility can be established and the optimisation is always feasible if it is feasible at the beginning.
 
\emph{Item 2)}. Now we establish the stability of CMPC with the help of  the recursive feasibility established above. 

There are two iterative processes involved in the above MPC algorithm: the current state and the terminal set. Following the discussion above, the terminal set shrinks with time since $\alpha(k+1)<\alpha(k)$. Once the terminal set approaches to zero, the update of the terminal set in Step 3 is terminated.  Then Lemma~\ref{thm:noT} can be used to show that the system state converges to the origin finally. Essentially, condition (\ref{eq:staConNoT}) is satisfied under CMPC. This is because by choosing $u(k+N|k)=0$ with the terminal state $x^*(k+N|k)=0$, the corresponding stage cost $l(x(k+N+1|k),u(k+N|k))=0$ (Assumption 1). Condition (\ref{eq:staConNoT}) is met since the stage cost $l(x^*(k+1|k),u^*(k|k))$ for any no-zero $x(k)$ is positive by the virtue of OSVF being a CLF (see Proposition 1 and Assumption 2).




Now we consider the case at the beginning where the initial state $x_o$ may be within the initial feasible terminal set $\Omega(\alpha_o)$ where $\alpha_o$ could be used to define a maximal control invariant set such that  OSVF is a CLF. This could happen, for example, when OSVF is a global CLF. In this case, when choosing $\alpha(0)=\alpha_o$, OSVF $m(x(k+N))$ associated with the terminal set $\Omega(\alpha(0))$ may be larger than OSVF $m(x(k))$ associated with $x_o$ so condition (\ref{eq:staConNoT}) is not necessarily satisfied. This can be fixed by recalculating $\alpha(0)$ as $\alpha(0)=m(x_o)-\delta$ with a sufficiently small $\delta>0$. By the virtue of $m(x))$ being a local CLF with respect to the updated $\Omega(\alpha(0)) \subset \Omega(\alpha_o)$, the optimisation is feasible and, in a similar fashion as previously, we can show that condition (\ref{eq:staConNoT}) is satisfied under the optimal control sequence.  After the first iteration, Step 3 in CMPC ensures the system state $x(k)$ is always outside the terminal set $\Omega(\alpha(k))$ in the following iterations by the way of updating the terminal set $\Omega(\alpha(k))$ 
\end{proof}


\begin{rem}
Minimisation is involved in checking the terminal constraint (\ref{eq:MPCS-NTC}), which may significantly increase  computational load. This can be avoided by reformulating the optimisation problem in Step 2. It can be shown that the terminal constraint (\ref{eq:MPCS-NTC}) is equivalent to that there exists $u(k+N) \in \mathcal{U}$ such that   
\begin{equation} \label{eq:MPCS-NTCN}
    l(x(k+N+1|k),u(k+N|k)) \leq \alpha (k) 
\end{equation}
This is because if the constraint $m(x(k+N+1|k)) \in \Omega(\alpha(k))$, i.e. Eq.(\ref{eq:MPCS-NTC}), is satisfied, one has 
\begin{equation}
    l(x^{*,1}(k+N+1|k),u^{*,1}(k+N|k))=m(x(k+N|k)) \le \alpha(k)
\end{equation}
with $x^{*,1}(k+N+1|k)=f(x(k+N|k),u^{*,1}(k+N|k))$ under the optimal control for OSVF, $u^{*,1}(k+N|k) \in \mathcal{U}$.  Therefore, there exists  $u(k+N|k) \in \mathcal{U}$ satisfying (\ref{eq:MPCS-NTCN}). Conversely, if there exists  $u(k+N|k) \in \mathcal{U}$ satisfying (\ref{eq:MPCS-NTCN}), one has 
\begin{equation}
    m(x(k+N|k)) \le l(x(k+N+1|k),u(k+N|k)) \leq \alpha (k) 
\end{equation}
which implies condition (\ref{eq:MPCS-NTC}) is satisfied. 

Hence the optimisation problem in Step 2 of CMPC can be re-written as 
\begin{equation} \label{eq:MPCSN=1}
     J^*(x(k))= \min_{u(k|k), \ldots, u(k+N-1|k), u(k+N|k) \in \mathcal{U}} J(x(k),U(k|k))  
\end{equation}
subject to system dynamics (\ref{eq:sys}), state and control constraint, and the terminal constraint (\ref{eq:MPCS-NTCN}). 

There is one more element in the control sequence to be optimised (i.e. $u(k+N|k)$) but the minimisation involved in checking the terminal constraint is avoid. The computational load is significantly reduced.   


\end{rem}

\section{OSVF properties and estimation of terminal set} \label{sec:OSVF}

Assumption A2 on OSFV $m(x)$ plays a central role in establishing stability of finite horizon MPC in this paper. The terminal set is also defined based on OSVF. Understanding the property of this function is important in checking if it is a local CLF and in estimating a maximal terminal set. 

\subsection{Properties of OSVF} \label{sec:POSVF}

Assumption A2 consists of two conditions (\ref{eq:OSVF1}) and (\ref{eq:OSVF2}) for OSVF. In essence, Condition (\ref{eq:OSVF1}) requires the stage cost $l(x^+,u)$ has a lower bound and upper bound for any $x \in \mathcal{X}$ and $u \in \mathcal{U}$. 
In this paper, we impose the following assumption on the stage cost to satisfy condition (\ref{eq:OSVF1}) for OSVF in Assumption 2. 

\begin{prop} \label{prop}
For any $x \in \mathcal{X}$, assume that there exist $\beta_1,\beta_2 \in \mathcal{K}_\infty$ such that (i) there exists a $u \in \mathcal{U}$ satisfying $l(x^+,u) \le \beta_2(|x|)$ and (ii)  $l(x^+,u) \ge \beta_1(|x|)$ for any $u \in \mathcal{U}$. Then condition (\ref{eq:OSVF1}) for OSVF in Assumption A2 is satisfied. 
\end{prop}
\begin{proof}
It follows from  $l(x^+,u) \ge \beta_1(|x|)$ for all any $x \in \mathcal{X}$ and $u \in \mathcal{U}$ that this property also holds for $m(x)$ which corresponds to the stage cost under one-step optimal control, i.e. $m(x) \ge \beta_1(|x|)$ for any $x \in \mathcal{X}$. It also holds for $x \in \Omega(\alpha)$ since it is a subset of $\mathcal{X}$. Since there exist a control $u \in \mathcal{U}$ and  $\mathcal{K}_\infty$ function $\beta_2$ such that $l(x^+,u) \le \beta_2(|x|)$, the OSVF $m(x)$ must be also upper bounded as it is defined as $\inf_{ u \in \mathcal{U}} l(x^+,u)$ for any $x \in \mathcal{X}$, hence for any $x \in \Omega(\alpha)$.
\end{proof}

Basically, Proposition~\ref{prop} states Assumption A2 holds if stage cost $l(x^+,u)$ is positive definite and well defined (i.e. there is no state in $\mathcal{X}$ such that stage cost approaches to infinity under all admissible controls). Similar conditions on $l(x,u)$ are widely required for establishing stability of MPC. For example, Assumption 2.14 in \cite{RawMay09} requires $l'(x,u) \ge \beta(|x|)>0$ for all $x \in \mathcal{X}$ and any $u \in \mathcal{U}$. Proposition~\ref{prop} is slightly more restrictive than $l(x,u)$ being positive definite. 

To further illustrate this condition, consider an affine system $x^+=f_0(x)+g(x)u$ with a stage cost $l=(x^+)^TQx^+ +u^TRu$, $Q>0$, and $R>0$. The corresponding OSVF can be calculated as
\begin{equation}
    m(x)=f_0(x)^T(Q^{-1}+g(x)^TR^{-1}g(x))^{-1} f_0(x)
\end{equation}
It can be shown that $m(x)$ is always positive definite and satisfies condition (\ref{eq:OSVF1}) if $f_o(x) \neq 0$ for any $x \neq 0$. It is acknowledged that there is significant research in MPC such as Economic MPC (EMPC) in relaxing the condition of the lower bound on the stage cost; e.g. \cite{faulwasser2018economic}.

We now focus on condition (\ref{eq:OSVF2}) for OSVF. It is equivalent to checking if there exists a control $u \in \mathcal{U}$ such that for all $x \in \Omega(\alpha)$ 
\begin{equation}
    m(f(x,u))-m(x) \le -\beta_3(|x|)
\end{equation}
In general, this can be carried out by solving the following optimisation problem
\begin{equation}
    \min_{u \in \mathcal{U}} m(f(x,u)) -m(x)
\end{equation}
and then assessing if the minimum is negative. 

For a general nonlinear system, similar to other CLF based approaches, it is not easy to directly check this condition. In the following, we investigate this problem for a linear system and it can also be extended to a nonlinear system by linearising it around the origin.

Consider a linear system
\begin{equation} \label{eq:linea2}
    x^+=Ax+Bu
\end{equation}
and a quadratic stage cost
\begin{equation} \label{eq:quadra}
    l(x^+,u)=(x^+)^TQx^++u^TRu
\end{equation}
with $Q>0$ and $R>0$. 
In the absence of constraints, solving the one-step optimisation problem with $l$ in Eq.(\ref{eq:quadra}) gives the optimal one-step control $u^{*,1}=-\left (B^TQB+R \right)^{-1} B^TQAx $
and the corresponding OSVF  
\begin{equation} \label{eq:l*1}
    m(x) = x^TA^T(Q^{-1}+BR^{-1}B^T)^{-1}Ax 
\end{equation}

It follows from (\ref{eq:l*1}) that $m(x)$ is positive definite so condition (\ref{eq:OSVF1}) is satisfied. Furthermore, we have
\begin{equation}
    m(x^+)=(Ax+Bu)^T (A^T(Q^{-1}+BR^{-1}B^T)^{-1}A) (Ax+Bu)
\end{equation}
Therefore, following condition (\ref{eq:OSVF2}), OSVF is a CLF if there exists a control $u$ such that
\begin{eqnarray}
   & &  (Ax+Bu)^T (A^T(Q^{-1}+BR^{-1}B^T)^{-1}A) (Ax+Bu) \nonumber \\
    & & -x^TA^T(Q^{-1}+BR^{-1}B^T)^{-1}Ax  <0
\end{eqnarray}

Letting $u=Kx$, the above condition is met if there exists a $K$ satisfying the following LMI
\begin{equation} \label{eq:LMI}
    \left[ \begin{array}{cc}
   A^T(Q^{-1}+BR^{-1}B^T)^{-1}A   & (A+BK)^T A^T \\
     A(A+BK)   & Q^{-1}+BR^{-1}B^T
    \end{array} 
    \right] > 0
\end{equation}
\begin{prop} \label{cor:LMPC}
Consider a finite horizon MPC problem for a linear system (\ref{eq:linea2}) and with a quadratic cost (\ref{eq:quadra}). Its OSVF is a CLF if there exists a gain $K$ such that LMI condition (\ref{eq:LMI}) is satisfied.     
\end{prop}

This approach can be extended to a nonlinear system to check whether OSVF is a (local) CLF by following similar ideas as in \cite{RawMay09} or \cite{chen2003terminal}. After a nonlinear system is linearised around the origin or approximated by a linear differential inclusion (LDI), the property of OSVF being a CLF for the linearised system is checked using condition (\ref{eq:LMI}). This gives a (local) CLF for nonlinear finite horizon MPC problems.

\subsection{An illustrative example} \label{sec:1storder}
We now illustrate the work so far using a simple example. It serves several purposes. First, it is used to provide insight how conservative the condition of OSVF being a CLF might be. Secondly, it illustrates how to use the developed condition to check the properties of an OSVF. Thirdly, it is used to compare the results in \cite{grune2008infinite} and \cite{chen2010stability} where stability of finite horizon MPC is analysed with different approaches.  

Consider an unconstrained first order linear system \citep{chen2010stability}
\begin{equation}
    x(k+1)=ax(k)+bu(k)
\end{equation}
and an one-step horizon cost function
\begin{equation} \label{eq:index2e}
    J(k)= qx(k+1|k)^2+ru(k|k)^2
\end{equation}
is employed with $q>0$ and $r > 0$. It is assumed that $b \neq 0$; otherwise, the MPC problem is ill-defined.  

A quick calculation gives OSVF as
\begin{equation}
 m(x)=\frac{a^2qr}{r+qb^2}x^2
\end{equation}

We show that this function is a CLF for the system with \emph{any} $a,b \neq 0,q>0,r>0$. For this MPC problem, condition (\ref{eq:LMI}) reduces to  
\begin{equation}
    \left[ \begin{array}{cc}
   \frac{a^2qr}{r+qb^2}      &  a(a+bK) \\
     a(a+bK)   & 1/q+b^2/r
    \end{array} 
    \right] > 0
\end{equation}

It can be easily shown that this condition is always satisfied, for example, choosing $K=-\frac{a}{b}$.

Therefore, OSVF for this finite horizon MPC problem is a CLF (globally in this case). We are able to construct a terminal set as in CMPC based on this property. By adding this terminal set into the original finite horizon MPC, CMPC in Section~\ref{sec:FHMPCS} is always feasible and, more importantly, it is (globally) asymptotically stable as stated in Theorem~\ref{thm:terminalset}. It shall be highlighted that \emph{CMPC is stable for any parameter $a, b \neq 0$, any positive control and state weight with any horizon $N$}.

This system has been investigated in \cite{grune2008infinite} with $a=2$ and $b=1$, $q=0.5$, $r=0$ using the approximation approach. The required horizon estimated for a stabilising MPC is from 12 to 23, depending on the performance approximation accuracy. Our result shows that CMPC is stable with \emph{any} horizon for this parameter setting. With $r=0$, Corollary~\ref{cor:LMPC} is not directly applicable.
However, this can be relaxed by not using the matrix identify but directly using the first part of (\ref{eq:l*1}).   

This example was also studied in  \cite{chen2010stability}  and a stability condition was given by
\begin{equation} \label{eq:exacon1}
   \frac{a^2r}{r+b^2q}<1
\end{equation}
The corresponding MPC is stable if the system parameters $a$ and $b$ and the state and control weights $q>0,r>0$ satisfy Eq.(\ref{eq:exacon1}). It can be shown that the stability condition proposed in this paper is much less conservative than that in \cite{chen2010stability}. 

This simple example clearly demonstrates that stability of finite horizon MPC can be established for a much wider range of system parameters or choices of the weights in the cost function by using the property of OSVF being a CLF. CMPC can stabilise the system with almost any combination of system parameters and positive weights in the cost function for this simple first order example with \emph{any} length of horizon. 

\subsection{Estimation of the maximal terminal set}

Once OSVF is a CLF, that is, there  exists a gain $K$ such that condition (\ref{eq:LMI}) is satisfied, we are able to construct a contractive terminal set based on that. Naturally we are interested in finding a maximally admissible terminal set.

First the state and control constraints $x \in \mathcal{X}$ and $u \in \mathcal{U}$ can be represented or approximated by the constraints 
\begin{equation}
    C_ix+D_iu \le 1, i=1,\ldots,s
\end{equation}
The set $\mathcal{D}$ in the state space corresponding to the above constraints is defined by
\begin{equation}
    \mathcal{D} = \{x \in R^n: (C_i+D_iK)x \le 1, i=1, \ldots, s \}
\end{equation}

With OSVF $m(x)$ given by (\ref{eq:l*1}), the terminal set $\Omega(\alpha)$ in (\ref{eq:levelset}) reduces to an ellipsoid set as
\begin{equation}
    \Omega(\alpha)=\{x \in R^n: x^TA^T(Q^{-1}+BR^{-1}B^T)^{-1}Ax  \le \alpha\} 
\end{equation}
It is well defined since $m(x)$ is positive definite under the condition that it is a CLF. 

Furthermore, it can be shown that $\Omega(\alpha)$ is contained in the set $\mathcal{D}$ if and only if
\begin{equation}
    (C_i+D_iK)(A^T(Q^{-1}+BR^{-1}B^T)^{-1}A)^{-1} (C_i+D_iK)^T \le 1/\alpha 
\end{equation}
with $i=1, \ldots,s$.
This condition is equivalent to 
\begin{eqnarray}
  &&  \left[ \begin{array}{cc}
   \frac{1}{\alpha}  & C_i+D_iK \\
     (C_i+D_iK)^T   & A^T(Q^{-1}+BR^{-1}B^T)^{-1}A
    \end{array} 
    \right] \ge 0; \nonumber\\
    &&\qquad i=1,\ldots,s
  \end{eqnarray}

Estimating the maximally admissible terminal set can be done by finding a gain $K$ such that $\alpha$ is maximised. This can be formulated as an LMI optimisation problem.  
\begin{equation}
   \alpha'^*= \min_{K} \alpha'
\end{equation}
subject to (\ref{eq:LMI}) and 
\begin{equation}
    \left[ \begin{array}{cc}
   \alpha'  & C_i+D_iK \\
     (C_i+D_iK)^T   & A^T(Q^{-1}+BR^{-1}B^T)^{-1}A
    \end{array} 
    \right] \ge 0
  \end{equation}
After this optimisation problem is solved, the maximal terminal set $\Omega(\alpha_o)$ in (\ref{eq:levelset}) is estimated with $\alpha_o=1/\alpha'^*$.

\section{Autonomous agent with finite horizon optimisation} \label{sec:example}

Optimisation with a finite look ahead horizon is currently used widely for online planning in robotics and autonomous systems where a lower level controller is designed to control a mobile robot/vehicle to follow a planned path under robot dynamics, disturbance and uncertainty; e.g. \cite{bostrom2019informative}, \cite{wongpiromsarn2012receding}. In the planning level, a more detailed dynamic model is ignored. Consider an autonomous agent (a mobile robot, or a vehicle) whose behaviour was abstracted as 
\begin{equation} \label{eq:linear}
    x(k+1)=Ax(k)+Bu(k)
\end{equation}
 with $x=[x_1,x_2]^T$ is the position of the agent and $u \in R^1$ is the move.  This type of model has been widely used in informative path planning \citep{bostrom2019informative} (e.g. Simultaneous localization and mapping (SLAM) \citep{temeltas2008slam}, or environment monitoring \citep{hutchinson2018information}) where the agent needs to decide the next move in order to maximise a reward or minimising a cost function. This can be formulated in an MPC setting through updating the environment belief and then performing an online optimisation repeatedly to decide the best next move; for example, see \cite{hutchinson2019unmanned};\cite{chen2021dual}. 
 
Consider an autonomous mobile robot whose behaviour is abstracted as a linear system (\ref{eq:linear}) with
\begin{equation} \label{eq:para}
    A=\left[ \begin{array}{cc}
   0.7      &  0.1 \\
      0.8   & 0.6
    \end{array} 
    \right]; \qquad  B=\left[ \begin{array}{c}
    0.8  \\
    -0.5 
    \end{array}
    \right]
\end{equation}
and its task is simply to go back to home at the origin. 
At each step, we choose to take a move to make the system's distance at the next time to the home as small as possible, i.e. minimising the cost function
\begin{equation} \label{eq:distance}
    J(x(k))=x_1(k+1)^2+x_2(k+1)^2+0.01u(k)^2
\end{equation}

The open-loop system is stable in the sense that it naturally goes back to the home without any control, $u=0$, as shown in Fig 1. However, by minimising the cost function (\ref{eq:distance}), the corresponding MPC scheme actually drives the system moving away from the origin and becomes unstable (see Fig 1). 


Now we calculate OSVF and check its property. The corresponding OSVF (\ref{eq:l*1}) is given by 
\begin{eqnarray}
    m(x)&=&x^TA^T(Q^{-1}+BR^{-1}B^T)^{-1}Ax\nonumber\\
    &=& 
    x^T \left[ \begin{array}{cc}
   1.1016     &  0.5891 \\
    0.5891  & 0.3162
    \end{array} 
    \right] x   
 \end{eqnarray}

It  can be shown that there exists a control gain $K$ such that Eq.(\ref{eq:LMI}) holds; for example, with $K=[-1.7454   -0.6510]$. Hence, $m(x)$ is a CLF. In the absence of state and control constraints, the whole space is feasible so CMPC globally asymptotically stabilises the agent.  In the presence of constraints, the maximum terminal set could be estimated following the procedure in Section~\ref{sec:OSVF}. For example, with $|u| \le 10$, the maximal terminal set could be estimated as $\Omega(\alpha_o)$
with $\alpha_o=1.297$.
CMPC with the above terminal set is always feasible and asymptotically stable with respect to the origin (see Fig 1). That is, we can make the original unstable MPC algorithm being globally stable by simply adding a terminal set as constructed in this paper.

\begin{figure}
\begin{center}
\includegraphics[width=\columnwidth]{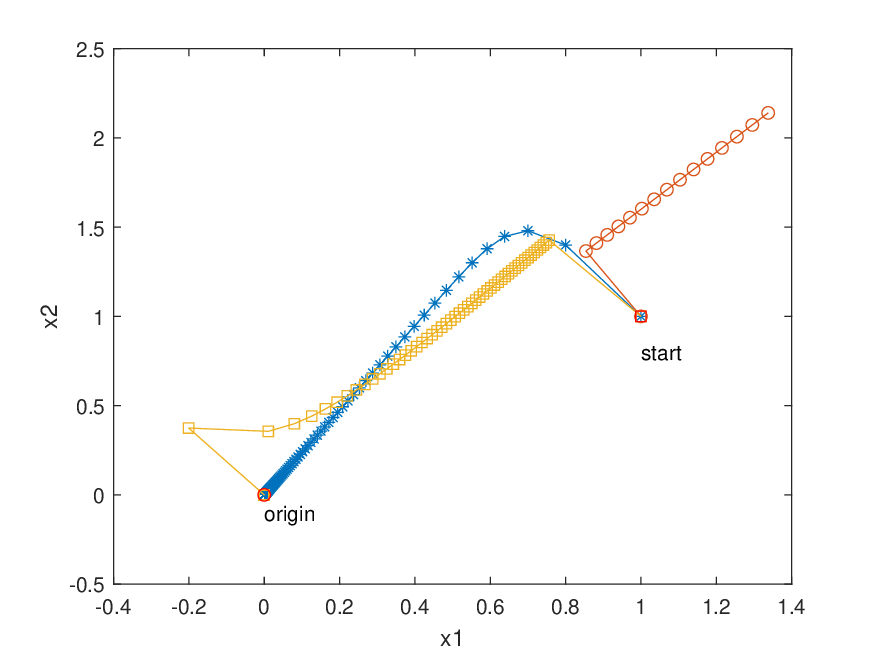}
\caption{Comparison of system trajectory under different control schemes. *: open loop; o: original MPC; square: CMPC}
\label{fig:state}
\end{center}
\end{figure}


\section{Conclusions} \label{sec:conclusion}

MPC algorithms with a limited horizon and without terminal weight are widely used in practical applications for many reasons. However, despite all the considerable effort and progress so far, it is still far way from proving a complete answer to stability analysis of this type of algorithms. Consequently, it still lacks of solid theoretic foundation particularly in stability analysis to underpin many their applications. The work in this paper contributes to bridging the gap in this area. By defining OSVF based on the stage cost, we show that stability of finite horizon MPC can be established if OSVF is a CLF. Under this condition, a contractive terminal set can be constructed and added to the original MPC. It shows that this slightly modified MPC algorithm, CMPC, is recursively feasible and asymptotically stable. 

Based on the work in this paper, a finite horizon MPC algorithm with stability guarantee can be developed by the following procedure. After defining the stage cost based on the performance requirement, we first check the property of OSVF. If it is a local CLF, we are able to establish stability of the finite horizon MPC algorithm using the proposed approach in this paper. If not, we may change the stage cost to make OSVF as a CLF subject to that the resultant performance meets the requirement. Otherwise, we resort to the existing solutions to ensure stability, e.g. adding a terminal weight, or increasing the horizon to be sufficiently large. 

Future research directions along the work proposed in this paper will be on reducing the conservativeness of the requirement of OSVF defined based on the modified stage cost being a clf, and extending the work to a wider class of MPC problems including MPC with terminal weight.  Furthermore, only regulation problems are considered in this paper and more research is required in extending the work to a broader range of MPC problems such as tracking or economic MPC.  It is anticipated that new complementary stability conditions will be developed for these MPC problems by exploiting the technique developed in this paper. The needs of less conservative and more effective stability analysis tools for MPC are not only due to their theoretical significance. They are also important in providing proven properties for practical applications, particularly for safety critical systems, extending the applicability of MPC, and providing more degrees of freedom for parameter tuning which potentially could improve performance. This work contributes to achieving a better trade-off between optimality in performance, stability, and computational burden.

\section*{Funding}
This work was supported by the UK Engineering and Physical Sciences Research Council (EPSRC) Established Career Fellowship (EP/T005734/1).

\section*{Disclosure statement}
No potential conflict of interest was reported by the author(s).

\section*{Data availability statement}
The datasets generated and(or) analysed during the current study are available from the corresponding author on reasonable request.

\section*{Notes on contributors}

\textbf{Wen-Hua Chen} holds a Chair in Autonomous Vehicles with the Department of Aeronautical and Automotive Engineering, Loughborough University, Loughborough, U.K., where he is leading the Centre of Autonomous Systems. He has a considerable experience in control, signal processing and artificial intelligence and their applications in robots, aerospace, and automotive systems. Prof Chen has contributed significantly to a number of areas including disturbance-observer based control, model predictive control and unmanned aerial vehicles. 
He is a Chartered Engineer, a Fellow of IEEE, the Institution of Mechanical Engineers and the Institution of Engineering and Technology, U.K. He has authored or coauthored near 300 papers and 2 books. Prof Chen currently holds the UK Engineering and Physical Science Research Council (EPSRC) Established Career Fellowship. 
\bibliographystyle{apacite}
\bibliography{common}
%

\end{document}